\documentclass[twocolumn]{aastex62}

\graphicspath{{./}{figures/}}

\shorttitle{Origin of the Black Hole Spin in Cygnus X-1}
\shortauthors{Qin et al.}
\begin{document}

\title{Hypercritical Accretion for Black Hole High Spin in Cygnus X-1}

\email{yingqin2013@hotmail.com}
\author{Ying\,Qin}
\affiliation{Department of Physics, Anhui Normal University, Wuhu, Anhui 241000, China}
\affiliation{Guangxi Key Laboratory for Relativistic Astrophysics, Nanning 530004, China}

\author{Xinwen\, Shu}
\affiliation{Department of Physics, Anhui Normal University, Wuhu, Anhui 241000, China}

\author{Shuangxi\, Yi}
\affiliation{School of Physics and Physical Engineering, Qufu Normal University, Qufu, 273165, China}

\author{Yuan-Zhu\, Wang}
\affiliation{Key Laboratory of Dark Matter and Space Astronomy, Purple Mountain Observatory, Chinese Academy of Sciences, Nanjing, 210033, China}



\begin{abstract}
Recent observations of AdLIGO and Virgo have shown that the spin measurements in binary black hole (BH) systems are typically small, which is consistent with the predictions by the classical isolated binary evolution channel. In this standard formation channel, the progenitor of the first-born BH is assumed to have efficient angular momentum transport. The BH spins in high-mass X-ray binaries (HMXBs), however, have been found consistently to be extremely high. In order to explain the high BH spins, the inefficient angular momentum transport inside the BH progenitor is required. This requirement, however, is incompatible with the current understanding of conventional efficient angular momentum transport mechanism. We find that this tension can be highly alleviated as long as the hypercritical accretion is allowed. We show that, for a case study of Cygnus X-1, the hypercritical accretion cannot only be a good solution for the inconsistent assumption upon the angular momentum transport within massive stars, but match its other properties reported recently.

\end{abstract}

\section{Introduction}
The recently revised measurements of the distance to Cygnus X-1 \citep{2021Sci...371.1046M} have shown that 
the masses of the black hole (BH) and its companion star are now significantly more massive than previous measurements,
 i.e., $M_{\rm BH} = 21.2_{-2.3}^{+2.2}$ $M_{\odot}$ and $M_1 = 40.6_{-7.1}^{+7.7}$ $M_{\odot}$. 
 To date the BH mass of Cygnus X-1 has exceeded the previously highest measured one in the X-ray binary for the extragalactic system M33 X-7. 
 We note that the BH dimensionless spin recently reported is extremely high, a$_*$ $>$ 0.9985 \citep{2021Sci...371.1046M}, 
 which is consistent with the previous measurement, namely $a_* >$ 0.983 \citep{2021ARA&A..59..117R,2014ApJ...790...29G,2014ApJ...788...93C,2011ApJ...742...84O,2021ApJ...908..117Z}.  
 In addition, it has been found that the surface helium-to-hydrogen ratio is about more than a factor of two relative to 
 the solar composition \citep{2012ARep...56..741S}.  
 Therefore, the observations of BH spin measurement and high surface abundance of its companion star has put a challenge for stellar models of massive binary evolution.

For Low-mass X-ray binaries (LMXBs), the BH spins span the entire range from zero to maximally spinning, which can be explained through the Eddington-limited accretion onto the BH after its birth \citep{2015ApJ...800...17F}. The BH spins, however, in the three HMXBs (Cygnus X-1, M33 X-7 and LMC X-1), have been found continuously to be spinning close to maximum. Such high spins cannot be explained when considering the limited lifetime of BH companion and the Eddington-limited accretion in the isolated binary evolutionary scenario.

\citet{2010Natur.468...77V} proposed a so-called Case-A mass transfer channel \citep{1967ZA.....65..251K} that is applicable to the formation of M33 X-7. In this channel, the two stars evolve initially in a close binary system, and the BH progenitor star, while still in its Main Sequence, initiates mass transfer onto its companion. \citet{2019ApJ...870L..18Q} systematically investigated this channel and found out that, in order to explain the three HMXBs, the inefficient angular momentum transport mechanism is required to form a fast spinning BH. 

To date, LIGO/Virgo have detected gravitational waves from $\sim$ 76 binary BH (BBH) mergers \citep{2021arXiv211103606T}. One of the most intriguing results is that the effective inspiral spins are typically low. This has been well explained in the classical isolated binary evolution channel \citep{2018A&A...616A..28Q,2020A&A...636A.104B,2020A&A...635A..97B}, in which the immediate progenitor of the BBH is a close binary system composed of a black hole and a helium star. In this classical channel, the second-born BH can be efficiently spun up by tides \citep{2018A&A...616A..28Q} from its companion and the first-born BH is assumed to have a negligible spin. This assumption requires that for a massive star the stellar core and its envelope have a strong coupling (i.e., efficient angular momentum transport inside stars), and thus the first-born BH would have a negligible spin as its progenitor evolves initially with very weak tides in a wide binary system and loses its envelope to its companion via  stellar winds and/or mass transfer.  Therefore, the current spin measurements of LIGO/Virgo are in favor of the efficient angular momentum transport mechanism. Such assumed mechanism, however, has a significant challenge when it is applied to the BH HMXBs.  

It is still unclear whether the angular momentum transport inside massive stars is efficient or not. The Tayler-Spruit dynamo \citep{1999A&A...349..189S,2002A&A...381..923S}, produced by differential rotation in the radiative layers, is considered as one of potential mechanisms responsible for the efficient transport of angular momentum between the stellar core and its radiative envelope. Stellar models with Tayler-Spruit dynamo (TS dynamo) cannot only reproduce the flat rotation profile of the Sun \citep{2005A&A...440L...9E}, but the observations for final rotation of neutron stars and white dwarfs \citep{2008A&A...481L..87S,2005ApJ...626..350H}. However, it has been recently found that models with TS dynamo still cannot explain the slow rotation rates of cores in red giants \citep{2014ApJ...788...93C,2012A&A...544L...4E}. More recently,  a revised TS dynamo \citep{2019MNRAS.485.3661F} was proposed to better match lower core rotation rates for subgiants, which is in a better agreement with asteroseismic measurements than predicted by the original TS dynamo.  But it was further confirmed that this revised TS dynamo still faces a challenge to reproduce the observed core rotation rates of red giant stars \citep{2019A&A...631L...6E}. To date a theoretical debate on the existence of the dynamo has been ongoing \citep{2007A&A...474..145Z}.

In the scenario of the classical binary evolution channel, the immediate progenitor of the BBH is a close binary system composed of a BH and a helium star. The first-born BH, formed from the more massive star, has been found with a negligible spin \citep{2018A&A...616A..28Q}. This result is exclusively dependent upon the well-accepted assumption of the TS dynamo for its progenitor. When considering the limited lifetime of HMXBs, breaking the Eddington accretion limit becomes a promising solution to explain the measured high BH spins. Early on, the case study of one HMXB \citep{2008ApJ...689L...9M} showed that the hypercritical accretion had happened in M33 X-7. \citet{2016MNRAS.459.3738I} argued that the BH accretion rate larger than 5000 times the Eddington limit is still stable.  Recently it was reported \citep{2020NewAR..8901542C} that SS433 is likely a BH X-ray binary, and that the inferred accretion rate is $\sim$ 10$^{-4}$ $M_{\odot}/yr$. The hypercritical accretions onto supermassive BHs in two-dimensional radiation hydrodynamical simulations have been performed \citep{2020ApJ...897..100V}. \citet{2021ApJ...912L..31W} pointed out that a BH accretion at a rate higher than the Eddington limit is well known and that it can also be another source of uncertainty affecting the theoretical estimates for the boundaries of pair-instability mass gap. Recent population studies \citet{2020MNRAS.497..302T} showed that the pair-instability mass gap might be polluted via either stable super-Eddington accretion or super-Eddington accretion in the Common Envelope phase.

The motivation of this work comes from the inconsistent BH spin measurements in two types of BH binaries (i.e., binary BHs and HMXBs). Such inconsistence has put different constraints on the efficiency of the angular momentum transport inside massive stars in the context of the classical isolated binary evolution channel \citep{2021arXiv210804821Q}. Additionally, the measured surface helium abundance of BH companion star is enhanced by more than a factor of two when compared with the solar composition. Combing the two together has put a significant challenge on the stellar models of massive binary evolution. Therefore, in this work under the assumption of non-spinning BHs at birth due to an efficient angular momentum transport inside massive stars, we study an alternative approach to forming fast-spinning BHs in HMXBs. In this study, we employ the the stellar structure code Modules for Experiments in Stellar Astrophysics \texttt{MESA} \citep{2011ApJS..192....3P,2013ApJS..208....4P,2015ApJS..220...15P,2018ApJS..234...34P,2019ApJS..243...10P} to investigate whether or not the hypercritical accretion can explain the currently reported high BH spin measurement and high surface helium abundance of BH companion of Cygnus X-1. The paper is organized as follows. In Section 2, we briefly introduce the hypercritical accretion of the BH. We then present in Section 3 the methods in this study. In Section 4, we show the result of the case study for Cygnus X-1 with the hypercritical accretion. Finally the discussion and conclusions are given in Section 5.

\section{Hypercritical accretion of a black hole}
% \label{sect:Hyper}
In this section we first present the accretion process onto a BH at the Eddington limit, and then briefly introduce the hypercritical accretion. The Eddington accretion rate is the maximum rate at which the outward force from the radiation pressure balances the inward gravitational pull. Considering a BH as the accreting object, its corresponding maximum accretion rate is defined as
\begin{equation}\label{m_edd1}
\dot{M}_{\rm edd} =  \frac{4\pi G M_{\rm BH}}{\kappa c \eta},
\end{equation}

where $\kappa$ is the opacity and it is assumed to be mainly due to  pure electron scattering, i.e., $\kappa$ = 0.2 (1 + X) $cm^2 g^{-1}$, $X$ is the hydrogen mass fraction, and $\eta$ the radiation efficiency. For $M_{\rm BH} < \sqrt{6} M_{\rm BH,init}$, $\eta$ is approximately expressed \citep{1970Natur.226...64B} as 
\begin{equation}\label{m_edd2}
\eta = 1 - \sqrt{1 - \left ( \frac{M_{\rm BH}}{3M_{\rm BH,init}} \right )^2},
\end{equation}
where $M_{\rm BH,init}$ is the initial mass of the BH before accretion. Under the assumption of the Eddington limit, the material in excess of the Eddington accretion rate is lost by carrying the specific orbital angular momentum of the BH. For an initially non-spinning BH, its mass and spin increase through accretion \citep{1970Natur.226...64B} according to 
\begin{equation}\label{m_edd3}
\dot{M}_{\rm BH} = (1- \eta) \dot{M}_{\rm acc},
\end{equation}

\begin{equation}\label{m_edd4}
a = \sqrt{\frac{2}{3}} \frac{M_{\rm BH,init}}{M_{\rm BH}}\left ( 4 -\sqrt{18\left ( \frac{M_{\rm BH,init}}{M_{\rm BH}} \right )^2 - 2} \right ).
\end{equation}
In our evolutionary sequences, none of the BH increases its mass by a factor of $\sqrt{6}$, i.e., $M_{\rm BH} < \sqrt{6} M_{\rm BH,init}$.

In case of the hypercritical accretion, the general point proposed \citep{1994ApJ...436..843B} is that if the mass transfer rates exceed the Eddington limit, the excess accretion energy can be removed by means of neutrino pairs rather than photons. This thus allows the matter to be smoothly accreted onto the BH. 

The hypercritical accretion can reach a rate for $\dot{M}/\dot{M}_{\rm Edd}$ $\sim 10^3$ or even higher \citep{1994ApJ...436..843B}. The case study \citep{2008ApJ...689L...9M} indicated that the hypercritical accretion had happened to the M33 X-7 system in which the BH was spun up through the hypercritical accretion after its birth. In the investigation of a binary system consisted of $M_{\rm BH}$ = 12 $M_{\odot}$ orbiting its companion star with mass $M_2$ = 25 $M_{\odot}$ at an orbital period of 6.8 days, it was found \citep{2003MNRAS.341..385P} that the companion star initiated overflowing its Roche lobe at the end of Main-Sequence phase and the corresponding mass transfer rate could reach a peak of $\sim$ 4 $\times$ 10$^{-3}$ $M_{\odot}$ $yr^{-1}$ on the thermal time-scale of the envelope. 

In this work, we employ the detailed binary evolution code \texttt{MESA} to investigate the evolution of a HMXB-like that could resemble Cygnus X-1. In this investigation, we allow for the hypercritical accretion, and assume the conservative mass transfer in the binary system.

\section{Methods}
% \label{sect:methods}
We use release 15140 of the \texttt{MESA} stellar evolution code to perform all of the detailed binary evolution calculations in this work. We adopt a metallicity of Z = $Z_{\odot}$, where $Z_{\odot}$ = 0.0142 \citep{2009ARA&A..47..481A}. We model the convection energy transport using the standard mixing-length theory \citep{1958ZA.....46..108B} with a mixing-length parameter of $\alpha = 1.93$. We adopt the Ledoux criterion for the boundary of the convective zones and choose the step core overshooting with the parameter $\alpha_{\rm ov} = 0.1$. We also adopt the convective premixing scheme as introduced in \citet{2019ApJS..243...10P} and include the thermohaline mixing with the parameter $\alpha_{\rm th}$ = 1.0. For superadiabatic convection in radiation-dominated regions, we employ the MLT++ to help numerical convergence \citep{2013ApJS..208....4P}.

For stellar winds, we use the \texttt{``Dutch''} scheme for both \texttt{RGB} and \texttt{AGB} phase, as well as the cool and hot wind. We adopt the default \texttt{RGB\_to\_AGB\_to\_wind\_switch = 1d-4}, a scaling factor \texttt{Dutch\_scaling\_factor = 1.0}, as well as \texttt{cool\_wind\_full\_on\_T = 0.8d4} and \texttt{hot\_wind\_full\_on\_T = 1.2d4}.

We model the angular momentum transport and rotational mixing diffusive processes \citep{2000ApJ...544.1016H}, including the effects of Eddington-Sweet circulations, the Goldreich-Schubert-Fricke instability, as well as secular and dynamical shear mixing. We adopt diffusive element mixing from these processes with an efficiency parameter of $f_c = 1/30$ \citep{2000ApJ...544.1016H,1992A&A...253..173C}. For an efficient angular momentum transport mechanism, we use the Spruit-Tayler dynamo \citep{1999A&A...349..189S,2002A&A...381..923S}. Mass transfer is modeled following the Kolb scheme \citep{1990A&A...236..385K} and the implicit mass transfer method \citep{2015ApJS..220...15P} is adopted. The timescale for orbital synchronisation is calculated following \citep{2002MNRAS.329..897H} for massive stars with radiative envelopes.

\section{Case study of Cygnus X-1}
\subsection{Updated properties of Cygnus X-1}
Cygnus X-1 is a binary consisting of a massive supergiant O-type star orbiting a BH with a 5.6-day orbital period. Recently the inferred BH and its companion masses of Cygnus X-1 with revised measurements of its distance have been reported to be more massive than previous measurements. The reported parameters with their median value and 68 per cent confidence interval boundaries of this system \citep{2021Sci...371.1046M} , are shown in Table~\ref{Tab1}.

\begin {table*}[htbp]
\caption {Main properties of Cygnus X-1}\label{Tab1}
\centering
% \scalebox{0.7}{
\setlength{\tabcolsep}{8.0mm}{
\begin{tabular}{ l  l  l  l}
\hline \hline
Parameters &Median &Lower bound &Upper bound \\
% \hline \\
$M_{\rm BH}/M_{\odot}$ &21.2 &18.9 &23.4 \\
$a_*$  &$>0.983$& & \\          
P/days &5.60 & &\\
$M_1/M_{\odot}$ &40.6 &33.5 &48.3\\
$\rm [He/H]$ &0.42 &0.37 &0.47\\
\hline 
\end{tabular}}
\end {table*}

\subsection{Application of the hypercritical accretion to Cygnus X-1}
\begin{figure*}[htp]
     \centering
     \includegraphics[angle=0,width=0.55\textwidth]{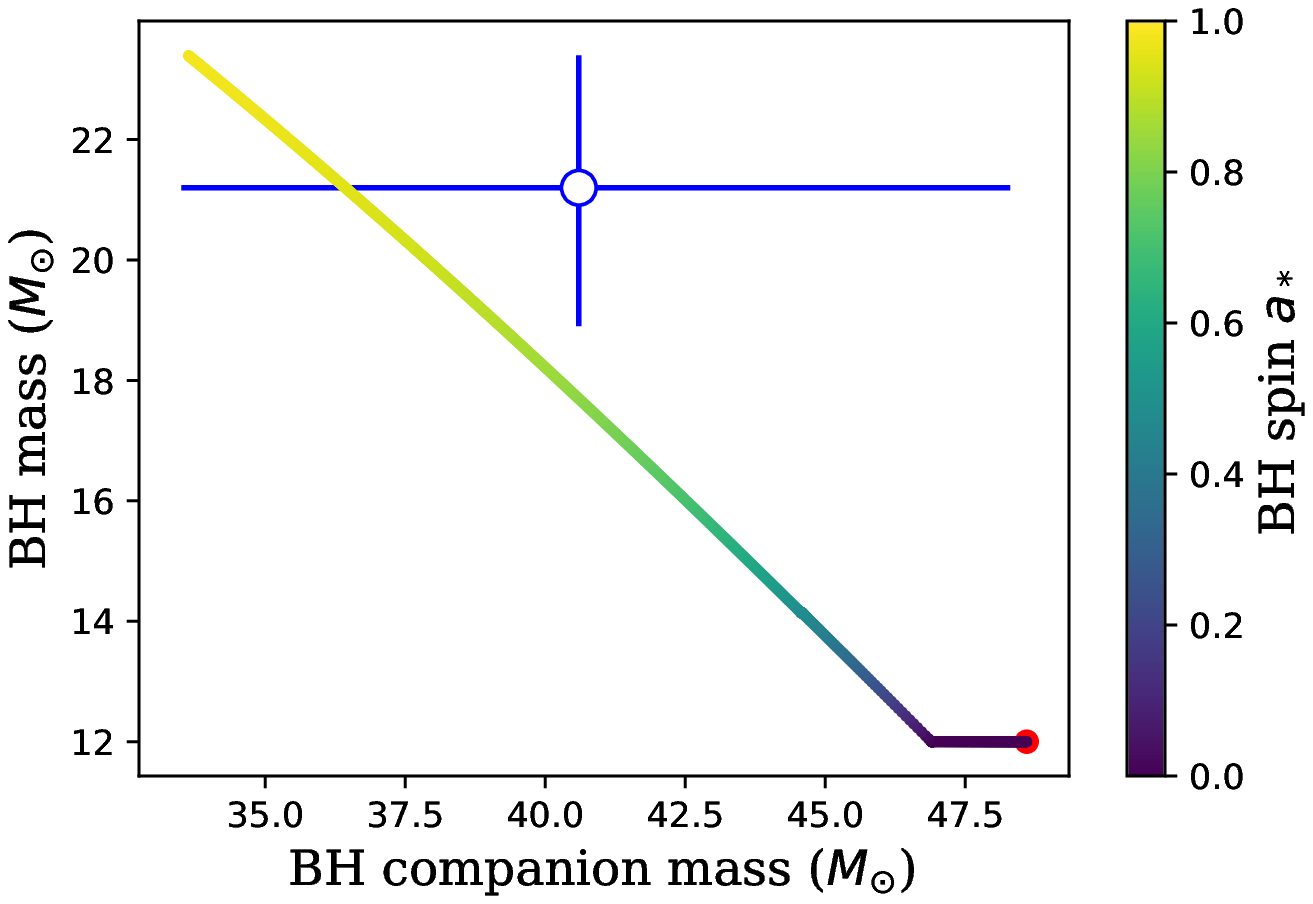}\includegraphics[angle=0,width=0.55\textwidth]{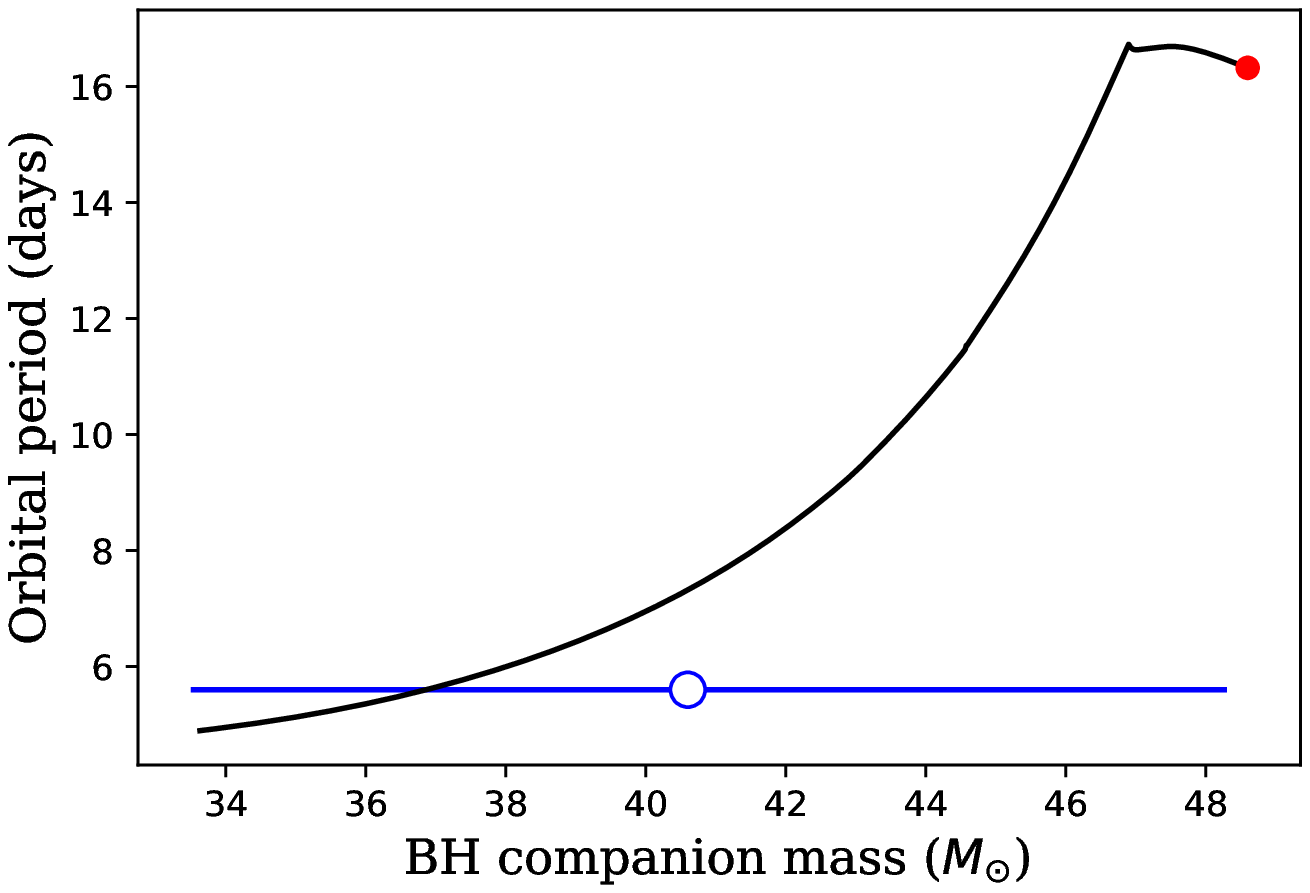}
     \includegraphics[angle=0,width=0.55\textwidth]{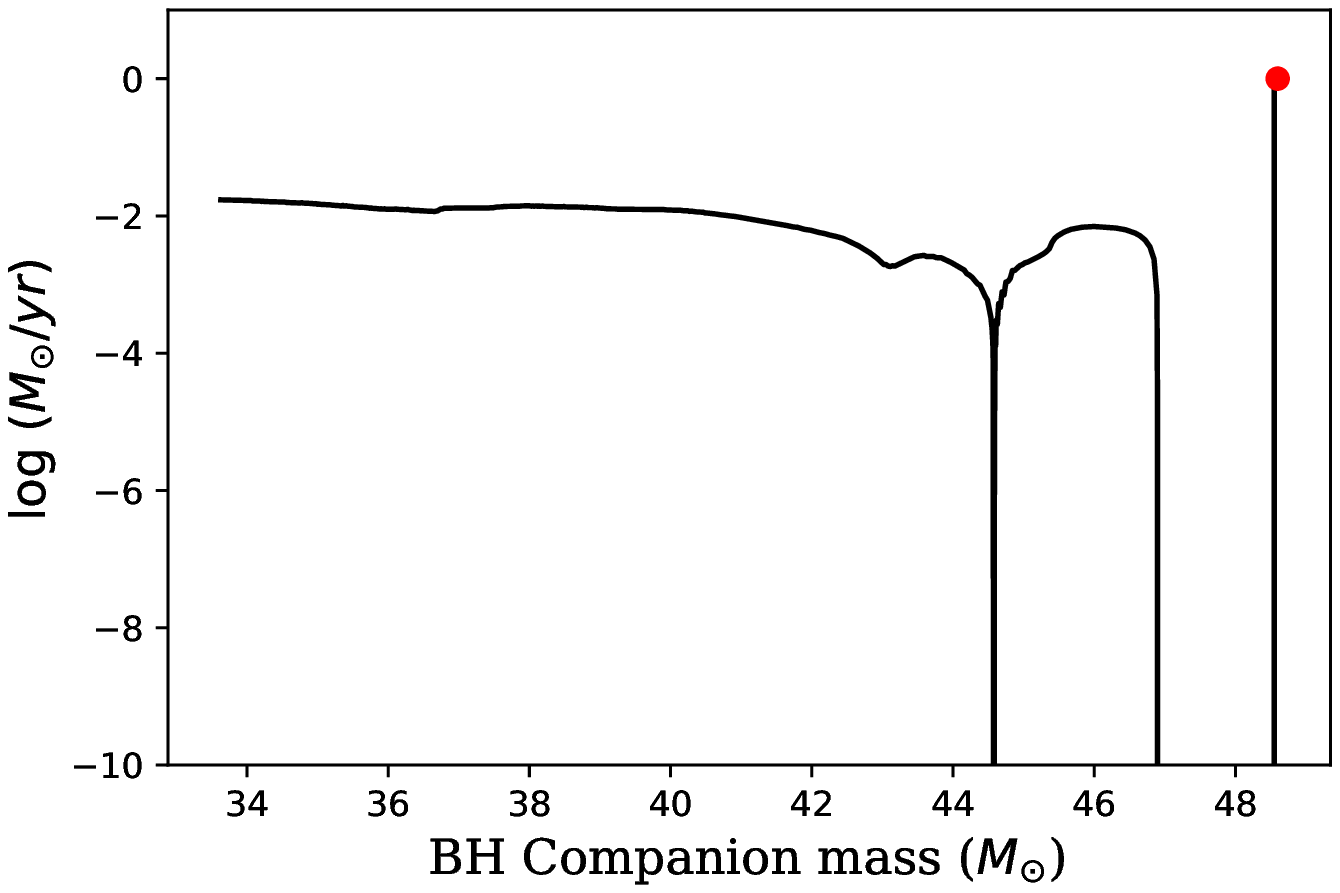}\includegraphics[angle=0,width=0.55\textwidth]{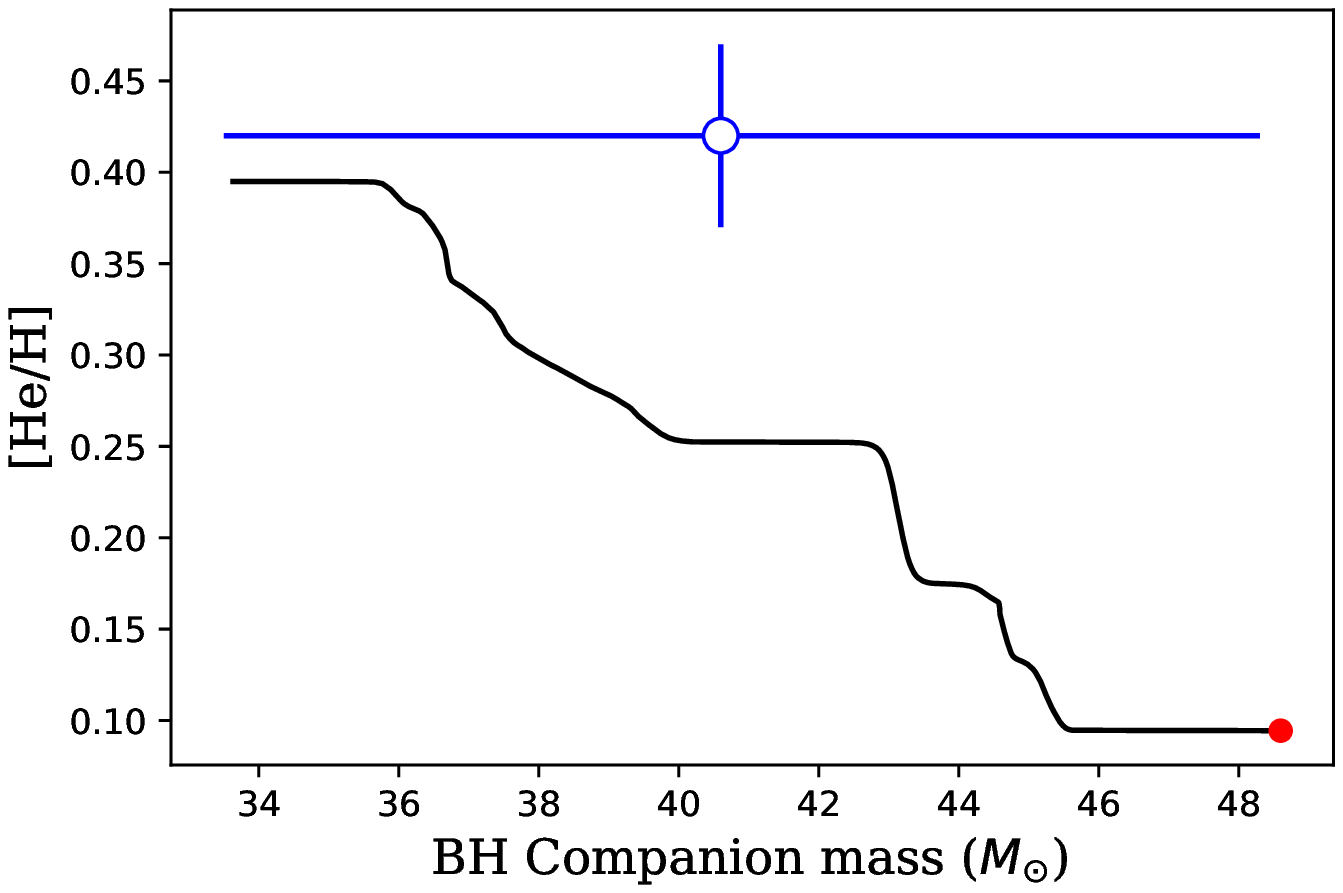}
     \caption{Key binary parameters as a function of the BH companion mass. Top left panel: the BH spin $a_*$ as a function of the BH mass and its companion star mass since the beginning of mass transfer (marked by a red point). The orbital period (top right panel), the mass transfer rate (bottom left panel), the ratio of the helium to hydrogen at the surface of the BH companion star (bottom right panel), as a function of the BH companion mass. The median values with corresponding 68 per cent confidence for observed properties are denoted in blue.}
     \label{cygnusx1}
\end{figure*}

Our result of the binary calculation that may resemble the formation history of Cygnus X-1 is shown in Fig.~\ref{cygnusx1}. We evolved a binary consisting of a BH with the mass $M_{BH}$ = 12 $M_{\odot}$ as a point mass and the companion star with its mass $M_2$ = 56 $M_{\odot}$ at zero-age Main Sequence, at an initial orbital period $P_{orb} = 13$ days.

In this numerical calculation, the BH had an assumption of zero spin at its birth. This is not only well accepted by currently conventional understanding for efficient angular momentum transport inside massive stars, but consistent with measured low BH spins from the gravitational-wave observations \citep{2021ApJ...913L...7A}. We assume that the material from the BH companion's winds captured by the BH is negligible when compared with the mass transfer through the Roche lobe overflow via the first Lagrangian point ($L_1$). In the top left panel in Fig.~\ref{cygnusx1}, we show the evolution of the binary after the onset of the mass transfer. It is shown that the BH gradually increases its spin magnitude as it accretes material from its companion star. The non-spinning BH accretes nearly half of its initial mass ($\sim$ 6 $M_{\odot}$) to reach a high spin close to maximum. The recently updated measurements \citep{2021Sci...371.1046M} of the BH mass, as well as its spin $a_*$ and companion mass, are marked in blue (at 68\% credibility).  The currently reported BH spin is very extreme, i.e., $a_* > $ 0.983, and it is still consistent with previous measurements \citep{2014ApJ...788...93C,2011ApJ...742...84O,2021ARA&A..59..117R}. Our finding shows that such high BH spin can be explained by the hypercritical accretion investigated here. We note that the current measurements of the masses for the BH and its companion have very larger uncertainties.

As the BH spin can be well explained by the hypercritical accretion, we then continue to show other parameters in our calculation. The top right panel in Fig.~\ref{cygnusx1} presents the evolution of the orbital period as a function of the BH companion mass. The binary orbital period first increases as the BH companion loses mass through the stellar winds. It then reaches the peak as the companion star expands to reach its Roche lobe. The mass transfer via the first $L_1$ from the BH companion (more massive) to the BH (less massive) shortens the orbital separation and thus the orbital period. In the bottom left panel, we present the mass transfer rate as a function of the BH companion mass. The gap shown after the first mass transfer phase is due to the quick shrink of the companion star. We note that the mass transfer is stable at a rate $\sim$ $10^{-2}$ $M_{\odot} /yr$. Such high value requires that the hypercritical accretion in this binary evolution is allowed. 

Furthermore, it was reported \citep{2012ARep...56..741S} that the abundance ratio of the surface helium-to-hydrogen is about a factor of two the solar composition. This indicates the BH companion has been stripped its outer hydrogen layer at a certain level. The bottom right in Fig. 1 clearly shows that this abnormality is reasonable due to its larger uncertainty of the measurements. Based on our investigation, such enhanced helium abundance is because that the BH companion star was exposed its inner layers through mass transfer and/or stellar winds. 

\section{Discussion and Conclusions}

%\st{This must have a red strikethrough line}

HMXBs are mostly considered as wind-fed binary systems \citep{2020ApJ...898..143S}, in which the BH is accreting part of the strong stellar winds of its companion. SS 433, known as a Galactic X-ray binary, is found that mass is lost from the system at a rate of $\sim 10^{-4}$ $M_{\odot}$ $yr^{-1}$, indicating that the compact object in SS433 is accreting mass from its companion at highly supercritical rate \citep{2004ASPRv..12....1F}, also see a for recent update \citep{2020NewAR..8901542C}. Analyzing the hard X-ray INTEGRAL observations of SS 433 provided reliable constraints on the binary mass ratio $\gtrsim$ 0.6, which suggests that the compact object is probably a BH. Assuming a BH as the compact object, the formation channel for SS 433 has been recently explored \citep{2020ApJ...896...34H}.  Additionally, the finding of the outflows \citep{2019A&A...623A..47W} for SS 433 indicates that the mass transfer onto the BH is nonconservative. However, as a case study for Cygnus X-1, we assumed that the mass transfer is conservative.  Note that the hypercritical rate here is significantly higher ($\sim$ two orders of magnitude) when compared with the above mass outflow rate.  Although this assumption is extreme, it will not significantly influence the final results since some fraction of the material can be significantly accreted by the BH in a supercritical accretion disk \citep{2004ASPRv..12....1F}. Therefore, our simplified assumption here is secure to test the efficiency for spinning up the accreting BH.

Recent population study shows the birthrate of Galactic BH binaries is a few $10^{-5}$ - $10^{-4}$  $yr^{-1}$ \citep{2019ApJ...885..151S}.  As only one such high-mass X-ray source in our Galaxy,  the formation rate of Cygnus X-1 is $\sim 10^{-3}$ $yr^{-1}$, which is challenging when considering our current understanding for massive binary evolution. M33 X-7 and LMC X-1,  known as another two BH HMXBs with high spin measurements, have been considered to share similar formation path. Early on, it was found \citep{2011MNRAS.413..183M} that, regardless of the formation channels for M33 X-7 and LMC X-1, the observed BH spins had to be obtained through the hypercritical accretion.  Additionally,  recent studies show that a BH can accrete materials from its companion via the stable mass transfer \citep{2017MNRAS.471.4256V,2021ApJ...920...81S}, but the hypercritical accretion is still required to efficiently spin up the BH when considering the limited lifetime of the companion star.

It has been suggested that the two types of BH binaries (BBHs and BH-HMXBs) likely have distinct formation paths \citep{2019IAUS..346..426Q,2021ARA&A..59..117R,2021arXiv211102935F}. This work is motivated by the inconsistent finding for the BH spin measurements in the two types of BH binaries. For BBHs measured from LIGO/Virgo \citep{2021ApJ...913L...7A}, the currently obtained low BH spins are in favor of the efficient angular momentum transport inside massive stars. On the other hand, in order to explain the high spin measurements for BHs in HMXBs, the angular momentum transport has to be inefficient \citep{2019ApJ...870L..18Q,2021arXiv210804821Q}. Given the classical isolated formation channel for the two BH binaries, this inconsistency has put a challenge on the angular momentum transport mechanism inside massive stars. This contradictory, however, can be alleviated as long as the hypercritical accretion is allowed for some cases, for instance Cygnus X-1, M33 X-7 and LMC X-1. 

Given the non-spinning BHs at birth, we then assumed in this work that the HMXB might have experienced the hypercritical accretion.  Therefore, we employ the detailed binary evolution code \texttt{MESA} to study the origin of the BH high spin for the HMXBs, specifically for the case of Cygnus X-1. We find that the binary evolution sequence shown in Fig.~\ref{cygnusx1} could resemble the Cygnus X-1 given its large uncertainties. In addition, the reported high ratio of the helium to hydrogen at the surface of the BH companion star can also be well explained. Given the very expensive computational cost, our study here is only the first step to investigate an alternative formation pathway, i.e., hypercritical accretion, for the case study of Cygnus X-1. As a follow-up work, we next plan to perform a more detailed investigation of the systematic parameter study for the three HMXBs (Cygnus X-1, M33 X-7 and LMC X-1).

\begin{acknowledgements}
The authors would like to thank Xiang-Dong Li for helpful discussions at the beginning of this project. Ying Qin acknowledges the support from the Doctoral research start-up funding of Anhui Normal University and the funding from Key Laboratory for Relativistic Astrophysics in Guangxi University. This work was supported by the National Natural Science Foundation of China (Grant Nos. 12192220, 12192221, U2038106) and by the Natural Science Foundation of Universities in Anhui Province (Grant No. KJ2021A0106).
\end{acknowledgements}

\bibliography{cyg}{}
\bibliographystyle{aasjournal}
\end{document}